%% file: main.tex
\documentclass[11pt]{article}

\usepackage{cite}
 \usepackage{subfigure}
\usepackage{multirow}
\usepackage{helvet}
\usepackage{amsmath}
\usepackage{amssymb}
\usepackage{setspace}
\usepackage{graphicx}
\usepackage{booktabs}
\usepackage{empheq}
\usepackage{soul}
\usepackage{tabularx}
\usepackage{array}
\usepackage{float}
\usepackage{braket}
\usepackage[utf8]{inputenc}
\usepackage{url}
\usepackage{hyperref}
\usepackage{slashed}
\usepackage[font=footnotesize,labelfont=bf]{caption}
\usepackage{color}
\usepackage{xspace}

\def\be{\begin{equation}}
\def\ee{\end{equation}}

\definecolor{darkgreen}{rgb}{0.0, 0.5, 0.13}

\newcommand{\fppdf}{\texttt{FPPDF}\xspace}

\usepackage[a4paper,top=3cm,bottom=3cm,left=2.5cm,right=2.5cm,bindingoffset=0mm]{geometry}

\begin{document}

\begin{center}


{\Large \bf Assessing the Impact of Fitting Methodology at aN$^3$LO with\\ \vspace{0.1cm}\fppdf : an Open Source Tool for Extracting Parton Distribution \\ \vspace{0.2cm} Functions in the Hessian Approach}

\vspace*{1cm}
J. M. Cruz-Martinez$^{a}$, T. Giani$^{b}$, L. A. Harland-Lang$^{c}$,\\    

\vspace*{0.5cm}
$^a$ Departamento de Física Atómica, Molecular y Nuclear, Facultad de Física, Universidad de Sevilla, E-41080 Sevilla, Spain \\  
$^b$ Dipartimento di Fisica, Universit\`a degli Studi di Torino and INFN, Sezione di Torino, Via Pietro Giuria 1, I-10125 Torino, Italy\\
$^c$ Department of Physics and Astronomy, University College London, London, WC1E 6BT, UK \\

\begin{abstract}
\noindent We present a new public code, \fppdf, to perform global fits of parton distribution functions (PDFs). The fitting methodology follows that  implemented by the MSHT collaboration, namely applying a fixed polynomial parameterisation of the PDFs and Hessian approach to error propagation, while for data and theory settings, the libraries used by the NNPDF collaboration are taken. This therefore complements the already publicly available NNPDF fitting code to enable fits with both neural network and fixed polynomial PDF parameterisations to be performed by the community, with otherwise identical theoretical and experimental inputs. As a first application, we use the new code to compare the PDFs found from fits at both NNLO and aN$^3$LO perturbative orders, but applying these two fitting approaches. We assess the impact of the two different methodologies on the PDFs and their uncertainties, providing results that complement previous comparisons between published PDF sets at NNLO and aN$^3$LO. We in particular find that the relative impact of going to the higher perturbative order and/or including missing higher order uncertainties is rather insensitive to which of these PDF parameterisation methodologies are used.

\end{abstract}

\end{center}
 
\begin{spacing}{1.2}
\clearpage
\tableofcontents
\clearpage
\end{spacing}

\section{Introduction}

Parton Distribution Functions (PDFs) play an essential role in the context of precision studies at current and future accelerator facilities. 
The increasing amount of precise data, the advances in the accuracy for theoretical prediction computations, and the new fitting methodologies are all leading to a significant reduction in PDF uncertainties, which in some cases are approaching the percent level.
Several groups  provide the community with independent PDF determinations based on different fitting methodologies and extended global datasets.
On the one hand, dedicated groups and experimental collaborations  perform PDF fits to subsets of data~\cite{H1:2015ubc,Alekhin:2017kpj,ATLAS:2021vod}, and on the other hand the  CT, MSHT and NNPDF collaborations perform global fits~\cite{Hou:2019efy,Bailey:2020ooq,NNPDF:2021njg}, which combine data from a range of experiments, with their results being combined in the past in the context of PDF4LHC~\cite{PDF4LHCWorkingGroup:2022cjn}.
New PDF sets with updated data and theory computations are regularly published, with the latest deliveries including a wealth of different experimental information from the  Large Hadron Collider (LHC) and beyond. The standard for theory predictions in global PDF determinations is now next-to-next-to-leading order (NNLO) precision in the QCD perturbative expansion, but a number of recent studies at approximate next-to-next-to-next-to-leading order (aN$^3$LO)  have now been performed, namely by the MSHT~\cite{McGowan:2022nag} and NNPDF~\cite{NNPDF:2024nan} collaborations.

With shrinking PDF uncertainties, non--negligible tensions between results from different groups have been observed, both in terms of the PDF central values and the size of the corresponding uncertainty.
A direct manifestation of this issue has been seen in a number of precision Standard Model measurements at the LHC, which can depend rather significantly on the input PDF~\cite{CMS:2024ony, ATLAS:2023fsi}. 
Several factors can be responsible for the discrepancies observed between independent PDF determinations. 
First, the differences in the datasets used in the fits and how are they treated. Second, the differences in some of the theoretical settings, such as the treatment of heavy quarks, for which the values of the heavy quark masses, variable flavour schemes used and most importantly treatment of the charm PDF itself can all differ; the latter is either independently parameterized and fitted (NNPDF) or computed using perturbative matching conditions (MSHT and CT). Third, the specific fitting parameterisation and approach to error propagation developed by different groups, with a neural-network (NN) representation for the PDF along with a Monte Carlo replica approach to error propagation applied by NNPDF, and a fixed polynomial parameterisation  with a Hessian approach to error propagation applied by MSHT and CT. With respect to the third point, it has been shown in Ref.~\cite{Harland-Lang:2024kvt} that, even when using exactly the same data and theory inputs, different methodologies can lead to results with non--negligible deviations from each other. 

In Ref.~\cite{Harland-Lang:2024kvt} fits to the NNPDF4.0 baseline dataset and theory settings were performed, using MSHT's (Chebyshev) polynomial parameterisation, and approach to error propagation. This could then be compared to the published NNPDF4.0 sets in order to gauge the difference in result due  to the PDF parameterisation alone. Differences between the two cases were observed for all PDF flavours, but particularly for individual quarks and anti--quark distributions, both in terms of central values and uncertainties. These results therefore highlighted the importance of understanding the impact different methodologies can have on the final PDFs. Along the same lines, recent studies have focused on the development of public frameworks to
perform PDF determinations using different statistical models, with the intent of assessing the impact of different methodological choices~\cite{Costantini:2025agd}.

The studies of Ref.~\cite{Harland-Lang:2024kvt} were performed using a privately developed fitting code, which made use of the publicly available NNPDF source code~\cite{NNPDF:2021uiq}. 
With the goal to facilitate and enable deeper investigation of these issues, and with the view to standardize these benchmarks between different approaches, in this paper we present a new public software: \fppdf.
This is based on the private code of Ref.~\cite{Harland-Lang:2024kvt}, and implements a Chebyshev polynomial basis for Hessian fits of PDFs, akin to the currently used MSHT methodology, but uses the public NNPDF libraries for data and theory computations.
The definition of the fit and preparation of the input is identical to the one used by the NNPDF framework and  runcards can be shared between both codes changing only settings which regard the fitting approach.
We thus provide a public framework that, combined with the already publicly available NNPDF code, can be used to reproduce the results presented in Ref.~\cite{Harland-Lang:2024kvt}, and further extend them. This tool allows any user to perform tests and studies to better understand the differences between the main methodologies used at the moment in PDF determinations, i.e. polynomial parameterisation with a Hessian approach to uncertainty propagation and neural-network parameterisation with a Monte Carlo approach to uncertainty propagation.
While other public tools, such as \texttt{xFitter}~\cite{Alekhin:2014irh} or the NNPDF framework~\cite{NNPDF:2021uiq}, permit PDF determinations with different theory settings or data selections, we focus here on the orthogonal scenario: PDF determinations with the exact same theory setting and data but different fitting methodologies.

As a first application of the new framework, we upgrade the study of Ref.~\cite{Harland-Lang:2024kvt} to  aN$^3$LO, producing a new family of Hessian sets at NNLO, NNLO+MHOU and aN$^3$LO accuracy.
We compare these Hessian PDF sets with analogous PDFs extracted in the NNPDF framework.
These two sets of results are based on the same theory grids for the cross sections and evolution,  treatment of heavy quarks, and  input datasets, in particular corresponding to those available in the public NNPDF code. Theory uncertainties due to missing and incomplete higher orders (denoted as MHOU and IHOU respectively) are also treated consistently in the two sets of PDFs, following the covariance matrix approach originally adopted by NNPDF~\cite{NNPDF:2024dpb}.
The only difference is therefore the fitting methodology used to parameterise the PDFs themselves\footnote{As discussed above there are other methodological differences, e.g. in the treatment of heavy flavours, but for brevity in this paper we will use the term `fitting methodology' to refer to those that concern the PDF parameterisation and error treatment.}.
By assessing the impact of N$^3$LO corrections when going from NNLO to aN$^3$LO PDFs in the two methodologies, we disentangle genuine N$^3$LO effects from discrepancies introduced by different methodological choices, and assess the impact of the different fitting frameworks for PDFs at up to aN$^3$LO accuracy.  We  find that the relative impact of going to the higher perturbative order and/or including missing higher order uncertainties is rather insensitive to which of these PDF parameterisation methodologies are used.

The outline of this paper is as follows. In Section~\ref{sec:fitting} we present a brief overview of the MSHT (Hessian) and NNPDF fitting methodologies to PDF fitting, focussing on the parameterisation, $\chi^2$ optimisation and approach to error propagation. In Section~\ref{sec:results} we present  our study of the impact of these fitting methodologies on fits at up to aN$^3$LO, providing results for both the PDFs and a selection of benchmark cross sections. In Section~\ref{sec:conc} we conclude. In Appendix~\ref{sec:app} we discuss the code usage and provide runcard examples.

\section{Overview of fitting methodologies}
\label{sec:fitting}

In this section we briefly describe the fitting methodology implemented in the new code. Given that in Sec.~\ref{sec:results}  we will compare our results with an analogous family of NNPDF sets, each point is supplemented with a brief description of the main differences with respect to the NNPDF methodological choices.
The discussion is kept minimal, and we refer for further technical details to the specific MSHT~\cite{Bailey:2020ooq} and NNPDF~\cite{NNPDF:2021njg} publications where the different methodological features are presented and discussed in detail.

\paragraph{PDF parameterization and fitting basis}

In  \fppdf, we take the exact same polynomial parameterisation used in the MSHT20 fit~\cite{Bailey:2020ooq}. To be precise, we take a parameterisation of the form
\be
xf(x,Q_0^2)~=~A(1-x)^\eta x^\delta \left( 1+\sum^n_{i=1} a_i T^{\rm Ch}_i(y(x)) \right),
\label{eq:1}
\ee
where $Q_0$ is the input scale, and $T^{\rm Ch}_i(y)$ are Chebyshev polynomials in $y$, with $y=1-2x^k$, with $k=0.5$. The PDFs are parameterised in the basis:
\[ u_V\,, d_V\,, S\,,s_+\,,s_-\,,\overline{d}/\overline{u}\,,g\,,c_+\,.\]
Here, $s_\pm = s\pm \overline{s}$, $S=2(\overline{u}+\overline{d}) + s_+$, and $c_+ = c+\overline{c}$ is freely parameterised if fitted charm is taken; both perturbative ($Q_0 = 1 \, \text{GeV}$) and fitted charm ($Q_0 = 1.65 \, \text{GeV}$ following NNPDF's settings) can be selected in \fppdf, although in the MSHT fits only perturbative charm is considered. The sum rules are imposed analytically, after which in the default case we are left with 52 free parameters at the fit stage. Further details can be found in Ref.~\cite{Bailey:2020ooq}.

In the NNPDF fit, the flavors which are independently parameterized are defined in the so--called evolution basis, maximally diagonalizing the DGLAP evolution equation and given by
\[ \Sigma\,, g\,, V\,,V_3\,,V_8\,,T_3\,,T_8\,,T_{15}\,.\]
The members of the fitting basis are parameterized by a single feed-forward neural network, supplemented by a polynomial pre--processing factor. 
Sum rules are imposed by computing numerically the integral of the relevant output nodes, and normalizing them to the desired values. In NNPDF4.0 this neural network has an 8-dimensional output for a total of $\sim800$ parameters per replica. We refer to Ref.~\cite{NNPDF:2021njg} for more details.

\paragraph{$\chi^2$ optimization}

In \fppdf, following the approach of MSHT, the central value PDF is found using the Levenberg–Marquardt minimisation algorithm~\cite{levenberg1944method,DonaldW:2006yco} to find the absolute minimum of the $\chi^{2}$, considering the entire dataset.

Instead, in the NNPDF approach, the parameters of each neural network are fitted independently by minimizing the $\chi^{2}$ of each data replica.
In addition, to avoid overffiting, a training/validation splitting procedure is implemented.
The central value PDF is then computed by averaging over a large number of replicas.
Note that the absolute minimum of the central data $\chi^2$ is therefore never expected to be reached: first because each replica is stopped as soon as its validation reaches a plateau, and second because the central data is never seen by the minimization algorithm.

Multiplicative uncertainties in the definition of $\chi^{2}$ are treated using a variant of the so-called $t_0$ method during the minimization~\cite{Ball:2009qv}, where in particular the covariance matrix definition is updated after every iteration of the Levenberg–Marquardt minimisation.

\paragraph{Error propagation and tolerance}

The number of free PDF parameters in the MSHT polynomial basis is sufficiently large to describe the data entering modern PDF fits, but remains small enough that we can effectively find a minimum in the PDF parameter space, and suitably expand around this after appropriate diagonalisation of the Hessian matrix. Having done this, we could define PDF uncertainties using a textbook $\Delta \chi^2=1$ criterion. This would be applicable to the ideal scenario of complete statistical compatibility between the multiple datasets entering the fit, a completely faithful evaluation of the experimental uncertainties within each dataset, including their Gaussian nature and correlations, and theoretical calculations that match these exactly. To account for deviations from this ideal scenario, in both the MSHT and CT fits an enlarged $\Delta \chi^2=T$ criterion is applied, effectively increasing the size of the PDF uncertainty with respect to the textbook case. A full discussion of these issues, and the precise method applied to calculate the `dynamic' tolerance in the MSHT fit can be found in Ref.~\cite{Martin:2009iq,Harland-Lang:2024kvt} (see also Ref.~\cite{Reader:2024cbb} for recent discussion on the contribution to this from experimental uncertainties).
In \fppdf both the dynamic and fixed (i.e. $T^2=1$ or some other global value)  options for the tolerance are available.

Given the high-dimensional parameter space characterizing the NNPDF parameterisation, the PDF uncertainty is defined by the ensemble of models that results from each independent minimization to a Monte Carlo replica of the data.
The resulting set of functions is taken to be representative of the underlying probability distribution in functional space.
We note that in a fixed polynomial basis, this method of error propagation has been shown to closely match the Hessian approach provided a $\Delta \chi^2=1$ criterion is applied~\cite{Watt:2012tq}.

\section{The impact of fitting methodologies at up to aN$^3$LO}
\label{sec:results}
In this section we use \fppdf to produce a family of PDF sets extracted using the Hessian formalism, and  compare them with analogous PDFs produced using the NNPDF methodology.
The analysis is done both at NNLO (reproducing some of the results of Ref.~\cite{Harland-Lang:2024kvt}), and aN$^3$LO. 
In order to better understand the features of the aN$^3$LO comparison presented here, we first provide a short description of the main differences observed between the two published aN$^3$LO sets, discussed in detail in the combination study of Ref.~\cite{Cridge:2024icl}. 
Having recalled the main conclusions of Ref.~\cite{Cridge:2024icl}, by inspecting the new set of results we are able to assess which differences are due to the methodologies and which ones are data/theory driven.

\subsection{Published aN$^3$LO sets}
As described in the introduction, two aN$^3$LO PDF determinations are currently available, by the MSHT~\cite{McGowan:2022nag} and NNPDF~\cite{NNPDF:2024nan} collaborations. 
The `a' denotes approximate, and refers to the fact that both sets do not achieve full N$^3$LO accuracy. The included N$^3$LO information is partial at two different levels: first because of the usage of approximate N$^3$LO splitting functions, reconstructed from their known small and large--$x$ behavior and from a finite set of Mellin moments; there is however now sufficient information to allow a rather precise extraction of the splitting functions.
Second, N$^3$LO corrections to hard cross sections entering in PDF determinations are only fully available for the massless Deep Inelastic Scattering structure functions, and not for fiducial cross sections for hadronic processes. 
Different strategies are used to account for the missing N$^3$LO information, as well as missing perturbative corrections at N$^4$LO and above: nuisance parameters for MSHT and theory covariance matrix for NNPDF.

The two published aN$^3$LO sets were first compared in Ref.~\cite{NNPDF:2024nan}. Their differences and similarities were further studied in Ref.~\cite{Cridge:2024icl}, where a combination of the two sets was presented. 
While most of the main differences were already present at NNLO, some additional discrepancies seem to emerge at N$^3$LO. This is especially true for the gluon PDF, which was found to differ more at aN$^3$LO than NNLO, with visible effects on the computation of processes such as Higgs production in gluon-gluon fusion. As discussed in Ref.~\cite{Cridge:2024icl}, this discrepancy can be only partially traced back to the different approximations used for $P_{gq}$: updating such approximations improves the comparison somewhat, with further updates with respect in particular to the transition matrix elements, as described in Ref.~\cite{Cridge:2025oel}, further improving the agreement. However, some non-negligible differences in the gluon still remain.

While some updates to the available aN$^3$LO information have appeared since the release of the public sets, Ref.~\cite{Cridge:2025oel} also shows that the resulting changes to the PDFs are only minor.
We therefore choose to focus on the information available at the time of Ref.~\cite{NNPDF:2024nan} since our goal is not to release a new PDF that supersede the previous releases from any of the two collaborations but rather compare the differences between them.

\subsection{Results}
In order to understand the origin of the observed differences at both NNLO and  aN$^3$LO, and to assess the impact of aN$^3$LO corrections, it is important to disentangle the effects due to data and theory inputs from those introduced by different methodological choices. 
In this section we achieve this by comparing the result of a Hessian fit produced using the \fppdf code with PDFs using the NNPDF methodology. The same input data and theory settings are used throughout, namely ones which closely correspond to the public aN$^3$LO NNPDF4.0 release.
In particular, in terms of the N$^3$LO information utilized, we are utilising the same evolved grids as in Ref.~\cite{NNPDF:2024nan}, which corresponds to \texttt{theoryID=40000000} at NNLO and \texttt{theoryID=722} at aN$^3$LO.
These settings correspond to a fitting scale of $Q_{0}=1.65$ GeV, $\alpha_{s}(m_{z})=0.118$ and the $\bar{\text{MS}}$ renormalization scheme. Additional technical parameters and theory settings follow exactly those of Ref.~\cite{NNPDF:2024nan}.
We will for concreteness refer to the first set of PDFs as fixed parameterisation fits, and the second as NNPDF fits.

\subsubsection{$\chi^2$ trends}

\begin{table}[!t]
  \scriptsize
  \centering
  \renewcommand{\arraystretch}{1.4}
  \input{tables/tab_chi2_process_msht}
  \vspace{0.3cm}
    \caption{The fit quality, $\chi^2$, for the different fixed parameterisation fits. The datasets are grouped according to the same process categorization as that used in Ref.~\cite{NNPDF:2024dpb} where the exact breakdown of datasets (as well as references to the experimental papers) can be found. In aN$^3$LO fits IHOU are always included.}
  \label{tab:chi2_byprocess_msht}
\end{table}

\begin{table}[!t]
  \scriptsize
  \centering
  \renewcommand{\arraystretch}{1.4}
  \input{tables/tab_chi2_process_nnpdf}
  \vspace{0.3cm}
  \caption{Same as in Tab.~\ref{tab:chi2_byprocess_msht} but for the NNPDF fits.}
  \label{tab:chi2_byprocess_nnpdf}
\end{table}

The $\chi^2$ values are reported in Tabs.~\ref{tab:chi2_byprocess_msht},~\ref{tab:chi2_byprocess_nnpdf} for the different fixed parameterisation and NNPDF fits respectively. 
Different datasets are grouped according to the same process categorization as in Ref.~\cite{NNPDF:2024dpb}.
A number of trends common to both methodologies can be observed:
\begin{itemize}
    \item For the NNLO results with and without MHOU, for almost all the processes the $\chi^2$ values slightly improve when adding MHOU, with the exception of top pair production, for which the $\chi^2$ increases  when including the theory covariance matrix, similar to the trends seen in Ref.~\cite{NNPDF:2024dpb}.
    \item A similar pattern is observed in the aN$^3$LO fits with and without MHOU, with the deterioration of the $\chi^2$ for top pairs data  more noticeable.
    \item When moving from NNLO+MHOU to aN$^3$LO+MHOU the fit quality remains roughly the same, in line with what was observed in Ref.~\cite{NNPDF:2024nan}.
    \item The fit quality in the fixed parameterisation case is somewhat lower in the no MHOU fits, consistently with the observation in Ref.~\cite{Harland-Lang:2024kvt}, but interestingly after including MHOUs little difference is observed. However, some care is needed when comparing the absolute numbers, due to the different optimization strategies\footnote{We recall in particular from Section~\ref{sec:fitting} that the NN optimization does not seek an absolute minimum of the $\chi^2$, while the Hessian optimization does.}
\end{itemize}
A key observation here is that the above trends are {\it not} particularly sensitive to the fitting methodology, even with the rather significant differences between the NNPDF and fixed parameterisation methodologies. This encouraging degree of stability indicates that the impact of the  aN$^3$LO corrections and missing higher order uncertainties is not particularly tied up with the PDF fitting methodology with respect the PDF parameterisation.

\subsubsection{Impact of aN$^3$LO corrections on NNLO PDF sets}
\begin{figure}
\begin{center}
\includegraphics[scale=0.4]{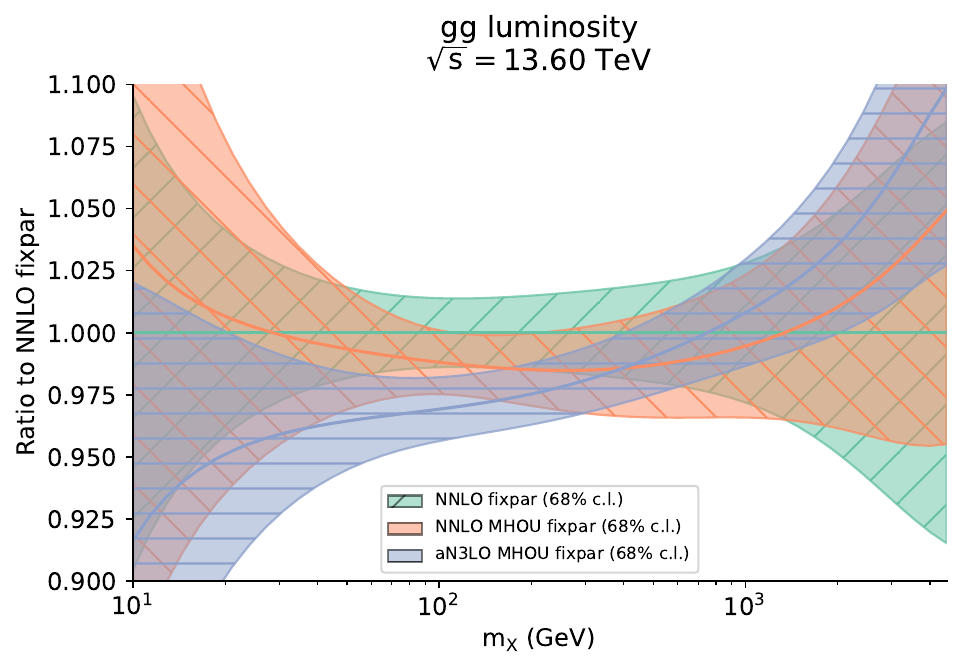}
\includegraphics[scale=0.4]{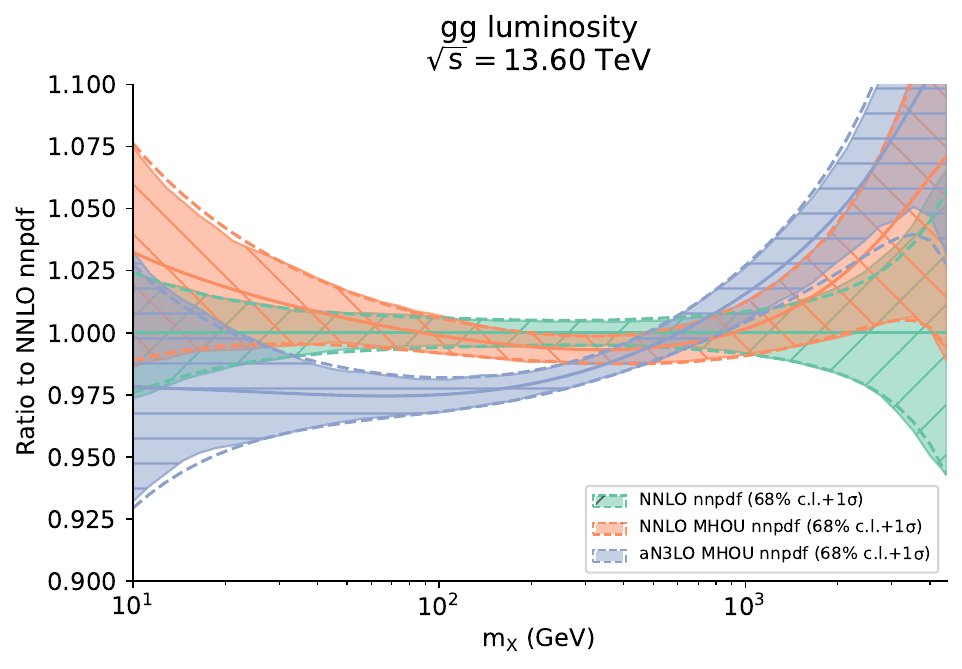}
\includegraphics[scale=0.4]{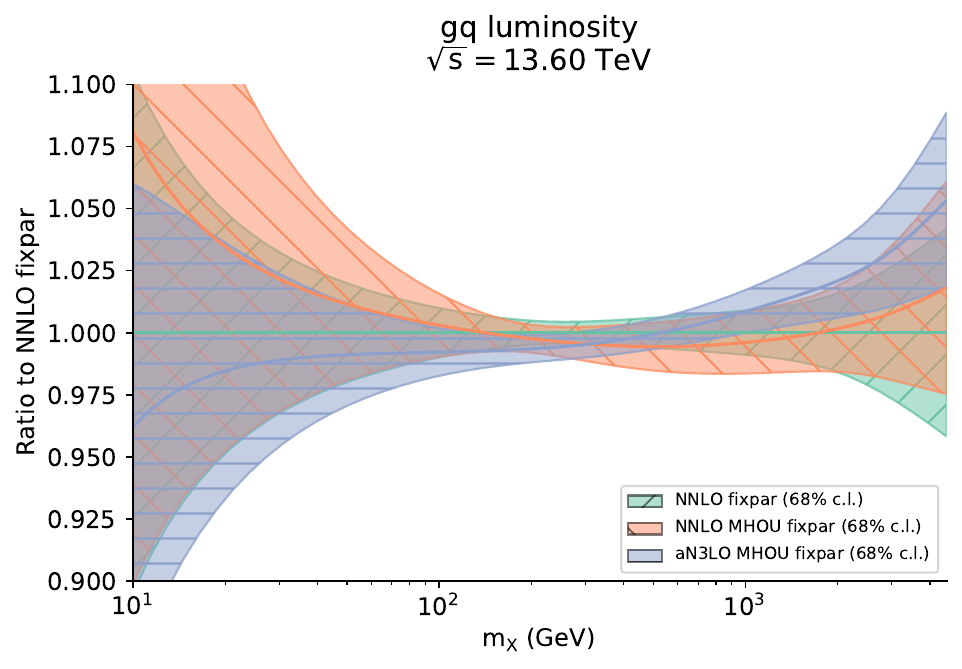}
\includegraphics[scale=0.4]{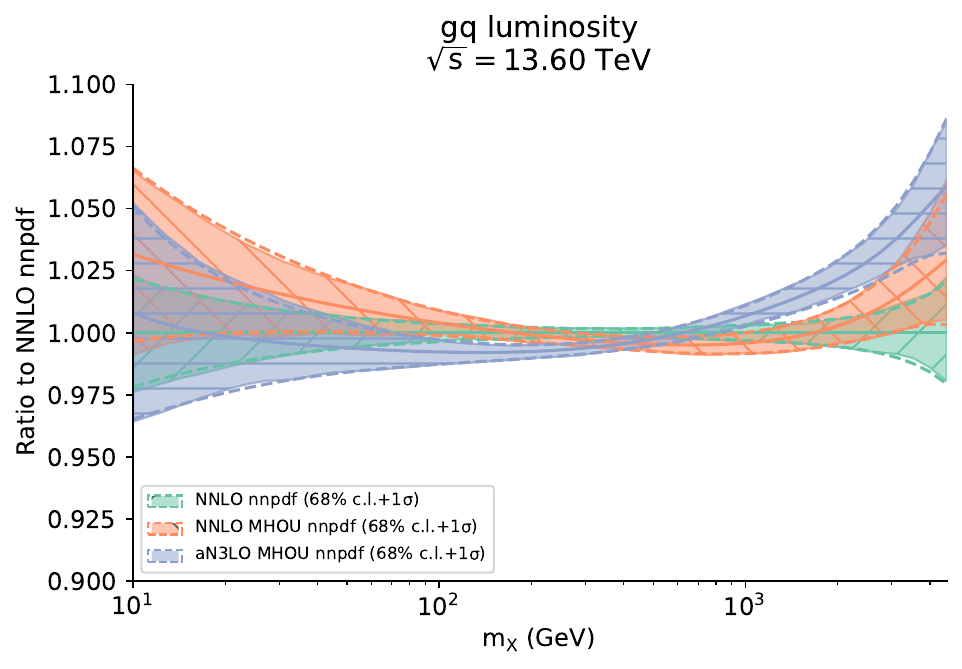}
\includegraphics[scale=0.4]{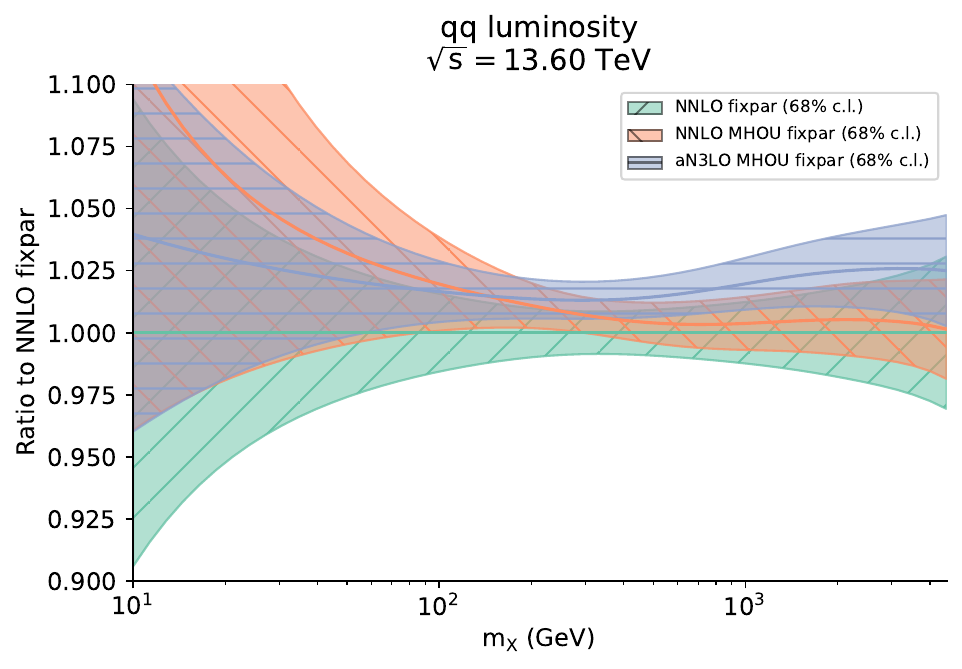}
\includegraphics[scale=0.4]{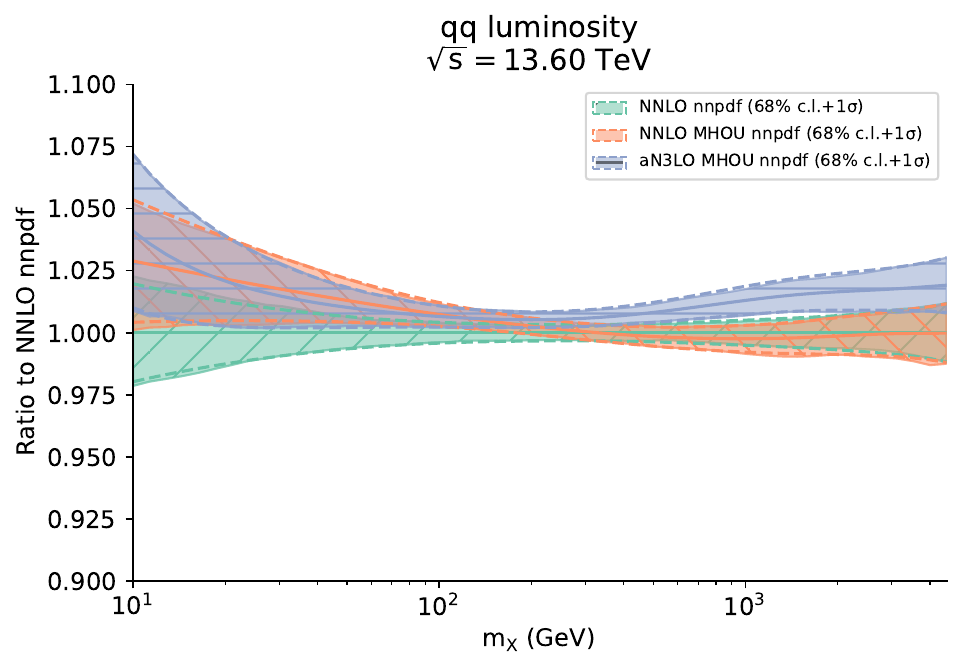}
\includegraphics[scale=0.4]{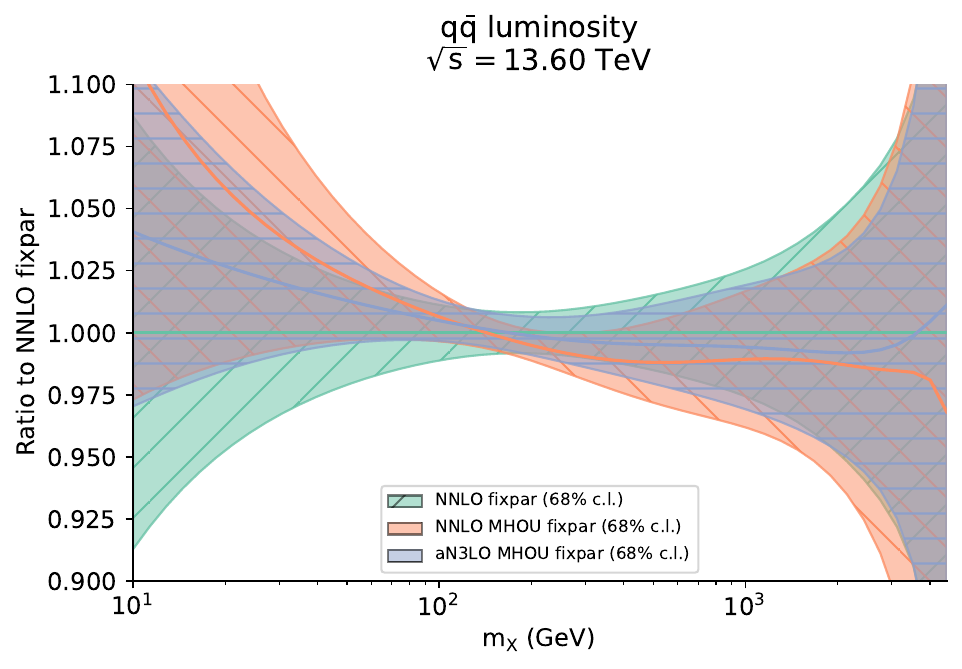}
\includegraphics[scale=0.4]{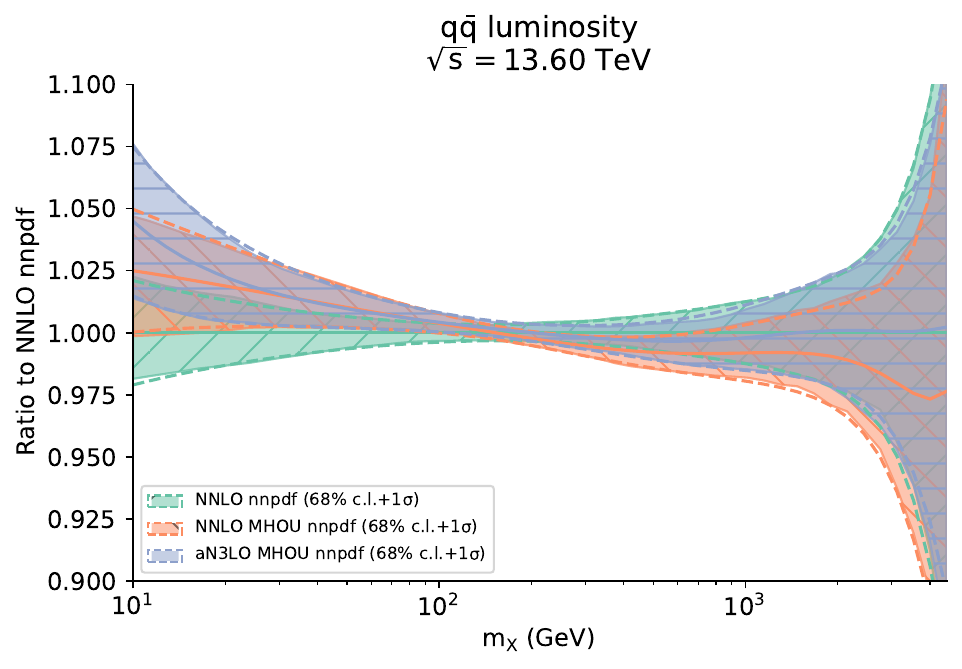}
\caption{\sf Parton luminosities at NNLO, NNLO+MHOU and aN$^3$LO+MHOU for the fixed parameterisation and NNPDF methodologies. Fixed parameterisation uncertainties are calculated using the dynamic tolerance criterion.}
\label{fig:aN3LO_lumi}
\end{center}
\end{figure}

We assess the impact of aN$^3$LO corrections in Fig.~\ref{fig:aN3LO_lumi}, where parton luminosities at NNLO, NNLO+MHOU and aN$^3$LO+MHOU accuracy are plotted for the fixed parameterisation (left panel) and NNPDF (right panel) fits. We note that here and throughout the fixed parameterisation uncertainties are calculated using the dynamic tolerance criterion.
Similar trends can be noticed for both methodologies: aN$^3$LO corrections induce a suppression in the gluon-gluon luminosity at small and intermediate invariant masses, $m_X$, and an enhancement at large $m_X$, with a similar pattern also visible for the gluon-quark channel; the quark-quark aN$^3$LO luminosity
is somewhat enhanced with respect to the NNLO one, and the quark-antiquark channel shows a mild enhancement and suppression at small and large $m_X$, respectively.
Crucially, we observe that these trends are common to both the fixed parameterisation and the NNPDF set of results when moving from NNLO to aN$^3$LO. This demonstrates that these  are  genuine effects, driven by the inclusion of N$^3$LO corrections, rather than being tied up with either fitting methodology.

\begin{figure}
\begin{center}
\includegraphics[scale=0.4]{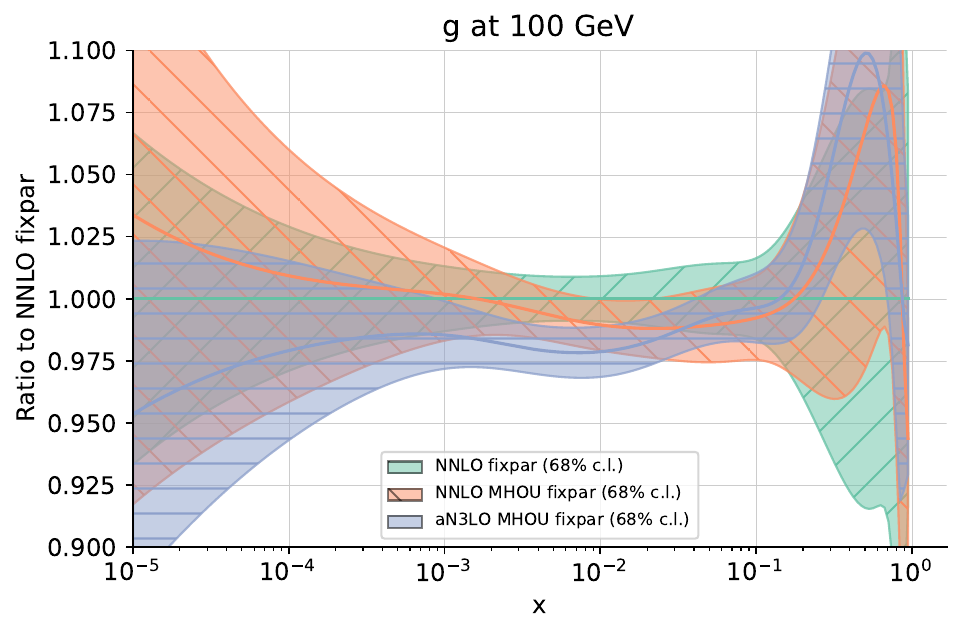}
\includegraphics[scale=0.4]{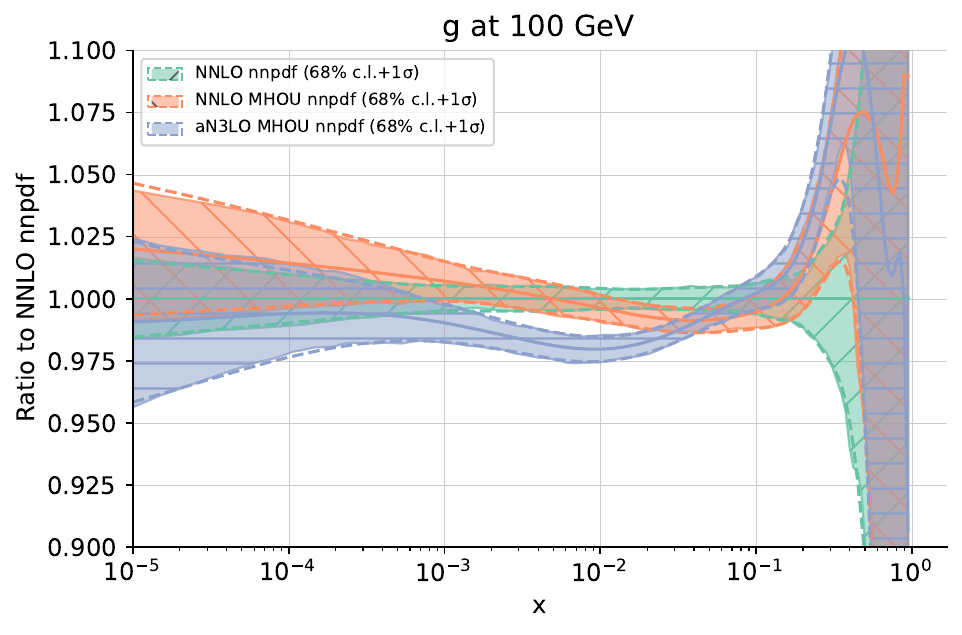}
\includegraphics[scale=0.4]{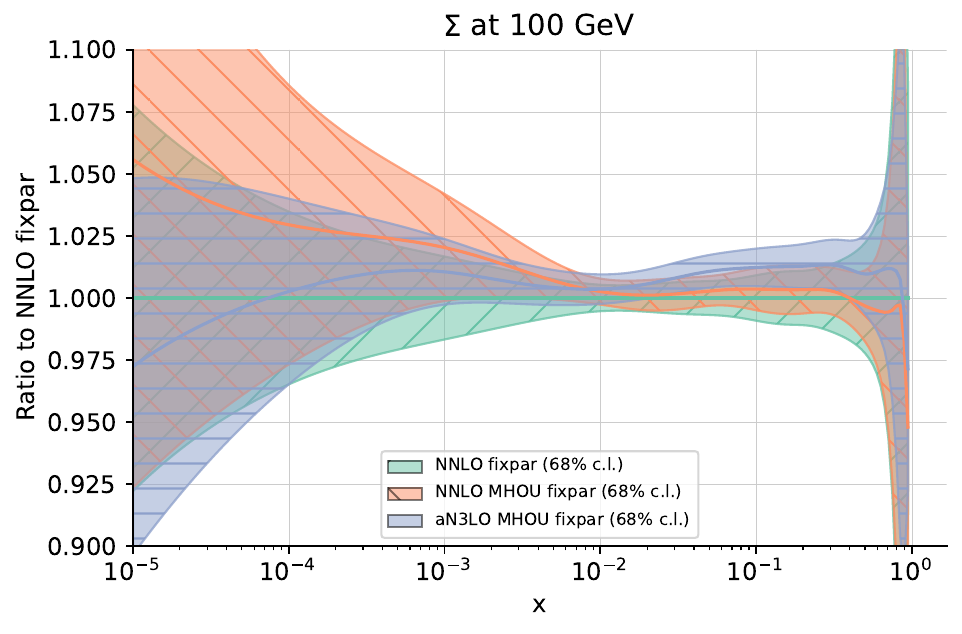}
\includegraphics[scale=0.4]{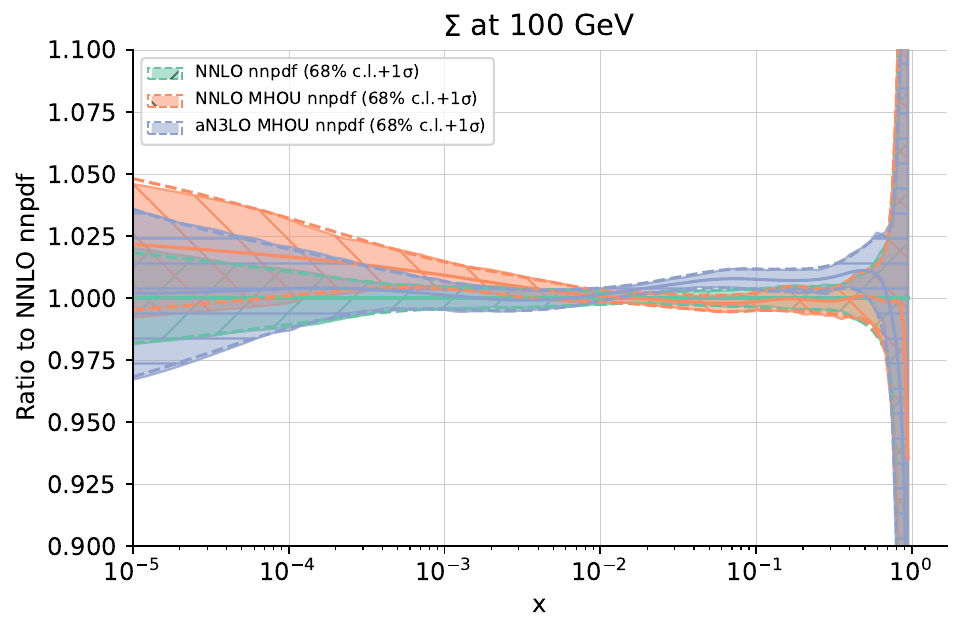}
\caption{\sf Gluon and singlet distributions at NNLO, NNLO+MHOU and aN$^3$LO+MHOU for the fixed parameterisation and NNPDF methodologies. Fixed parameterisation uncertainties are calculated using the dynamic tolerance criterion.}
\label{fig:aN3LO_gluon_singlet}
\end{center}
\end{figure}

Additional comparisons are shown at the PDF level in Fig.~\ref{fig:aN3LO_gluon_singlet}, where we plot the gluon and singlet distributions: the gluon is suppressed around $x\sim 10^{-2}$ and enhanced around $x \sim 0.5$, while the singlet shows a mild enhancement for $x \gtrsim 0.5 \times 10^{-2}$.
As in the case of the parton luminosities, the same general trends can be noticed for both the fixed parameterisation and NNPDF fits, when including N$^3$LO corrections.

\subsubsection{Comparison between aN$^3$LO results}

\begin{figure}
\begin{center}
\includegraphics[scale=0.4]{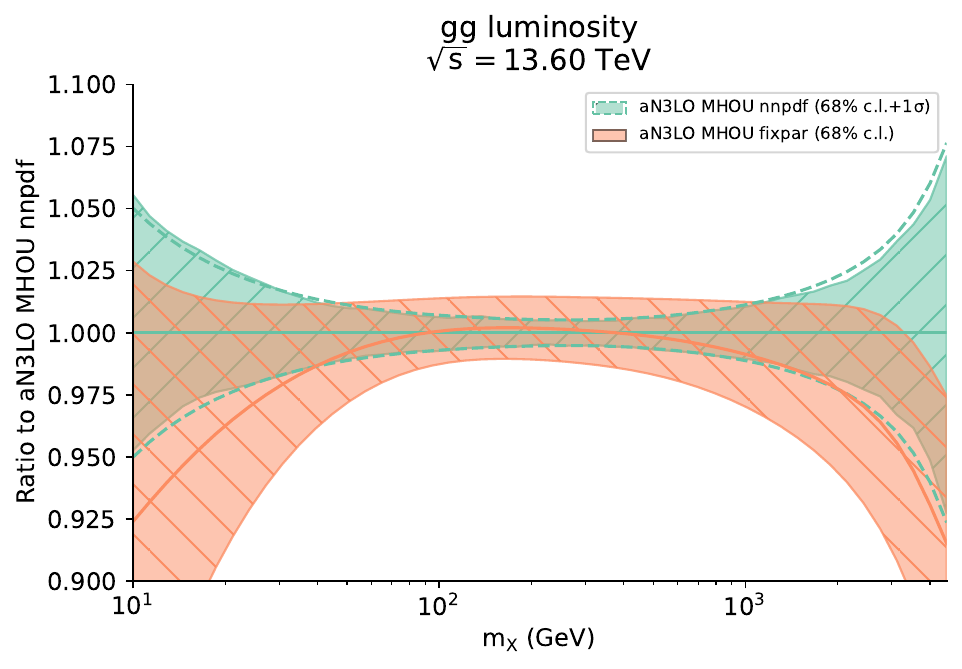}
\includegraphics[scale=0.4]{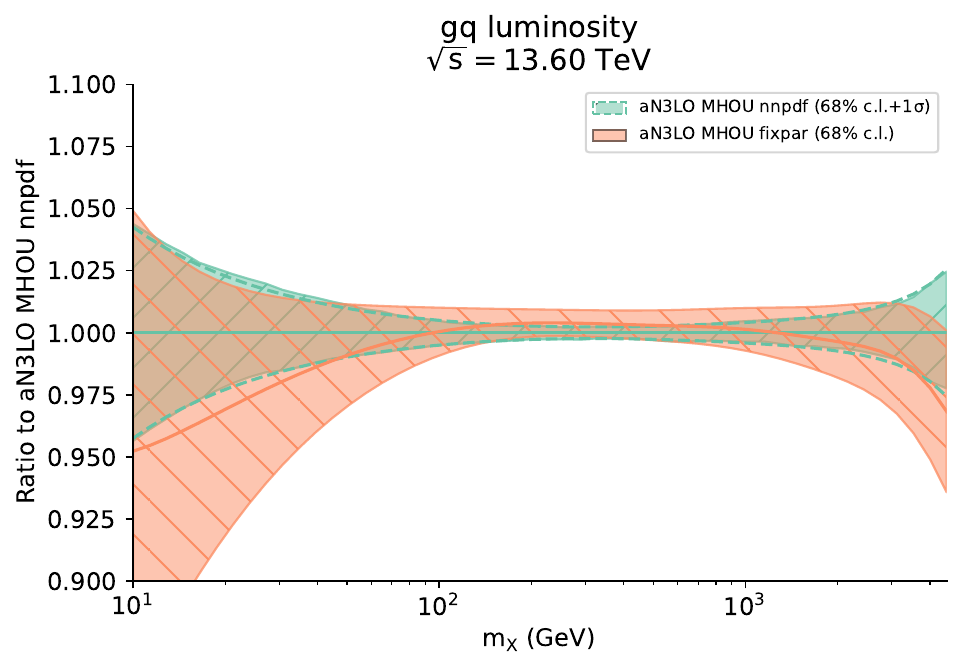}
\includegraphics[scale=0.4]{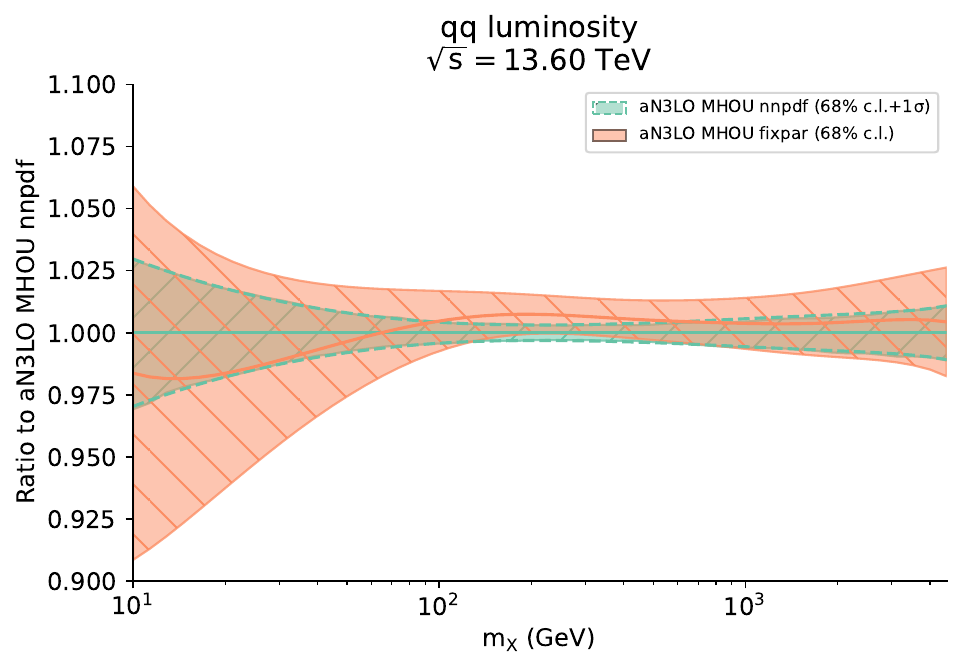}
\includegraphics[scale=0.4]{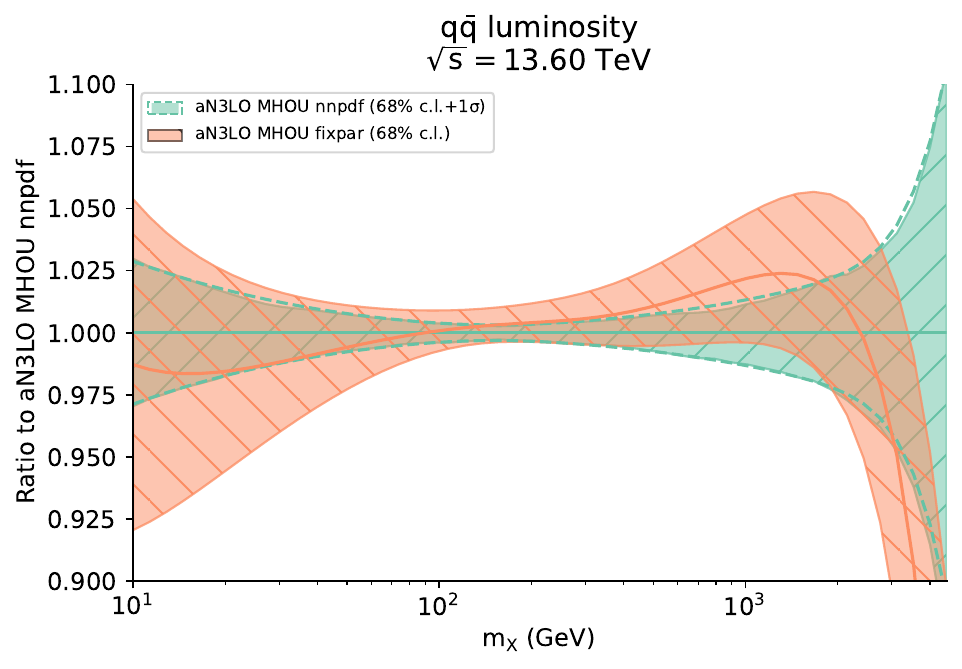}
\end{center}
\caption{\sf Comparison between parton luminosities at aN$^3$LO+MHOU for the fixed parameterisation and NNPDF methodologies. Fixed parameterisation uncertainties are calculated using the dynamic tolerance criterion.}
\label{fig:an3lo_comparison_lumi}
\end{figure}

\paragraph{Parton luminosities} In Fig.~\ref{fig:an3lo_comparison_lumi} we compare a selection of PDF luminosities at aN$^3$LO, for the fixed parameterisation and NNPDF fits.
We observe good compatibility between the two cases in the whole $m_X$ range: central values are always in agreement at 1$\sigma$ level, with the exception of the quark-antiquark channel for large invariant mass. 
We note how the same comparison between the published MSHT and NNPDF sets performed in Ref.~\cite{Cridge:2024icl} (see in particular Fig.~4) give somewhat different results, highlighting a tension around $m_X\sim 10^2$ GeV for the gluon-gluon and gluon-quark channels. 
Fig.~\ref{fig:an3lo_comparison_lumi} of this paper shows how such differences are due to the different treatment of data and theory settings and not to the specific fitting methodologies. A clear difference in the size of the uncertainties is also observed, with these being generally larger in the fixed parameterisation case. This will be commented on further below in a dedicated section, but in broad terms is due to the impact of the dynamic tolerance procedure in the fixed parameterisation fits, which roughly corresponds to taking a larger value $T \approx 3$ in the error definition, $\Delta \chi^2=T^2$.

\begin{figure}
\begin{center}
\includegraphics[scale=0.4]{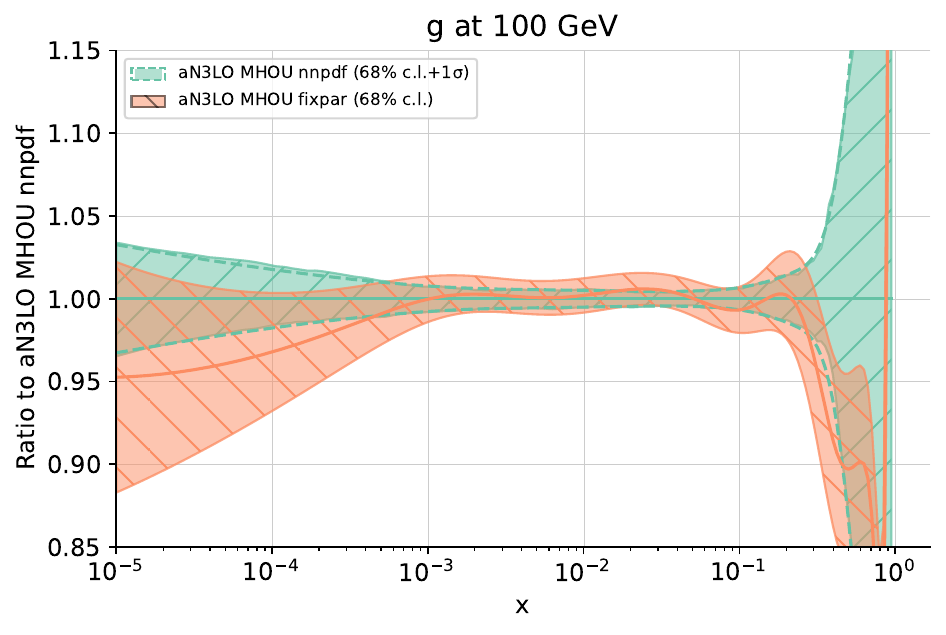}
\includegraphics[scale=0.4]{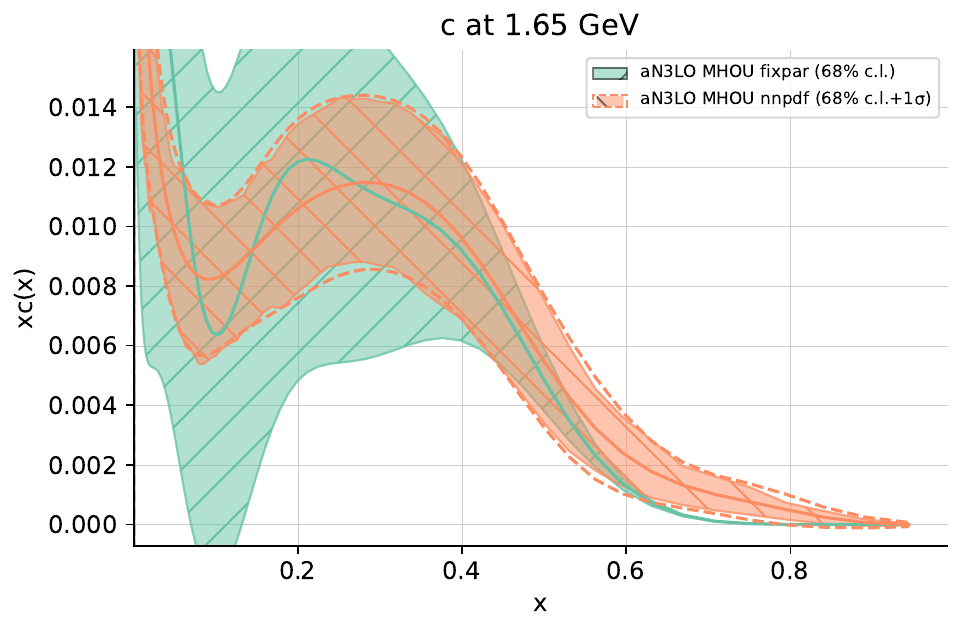}
\end{center}
\caption{\sf Left panel: comparison between the gluon PDF obtained using the fixed parameterisation and NNPDF methodologies. Right panel: comparison between the medium and large-$x$ charm at 1.65 GeV obtained using the MSHT and NNPDF methodlogies. Fixed parameterisation uncertainties are calculated using the dynamic tolerance criterion.}
\label{fig:gluon_charm_an3lo}
\end{figure}

\paragraph{Gluon PDF} A similar conclusion holds at the level of the gluon PDF, shown in  Fig.~\ref{fig:gluon_charm_an3lo} (left): while in the previous aN$^3$LO comparison of Ref.~\cite{Cridge:2024icl} some tensions were visible around $x\sim 0.5 \times 10^{-2}$, here we observe 1$\sigma$ compatibility for the whole $x$ range. 

\paragraph{Charm PDF} As discussed above, the published MSHT and NNPDF sets are based on different procedures to match massive and massless heavy quarks schemes. In addition in the NNPDF approach the charm PDF is independently parameterized and fitted to the data. These features can bring important differences not only in the charm PDF, but also in the behavior of the other flavors.  
Since the results presented in this section have been produced with identical NNPDF theory settings, it is interesting to compare the charm PDF obtained with the two methodologies. Charm quark PDFs from the fixed parameterisation and NNPDF methodologies are compared in the right panel of Fig.~\ref{fig:gluon_charm_an3lo}. The comparison is performed in linear scale at 1.65 GeV, given the interest of these settings for intrinsic charm studies. We observe 1$\sigma$ compatibility of the central values up to deep large-$x$ extrapolation, with the main differences being the size of the PDF error (see final paragraph of this section).
A comparison cannot be done using the published PDFs, given the lack of a parameterized charm in the released MSHT sets. 

\begin{figure}
\begin{center}
\includegraphics[scale=0.4]{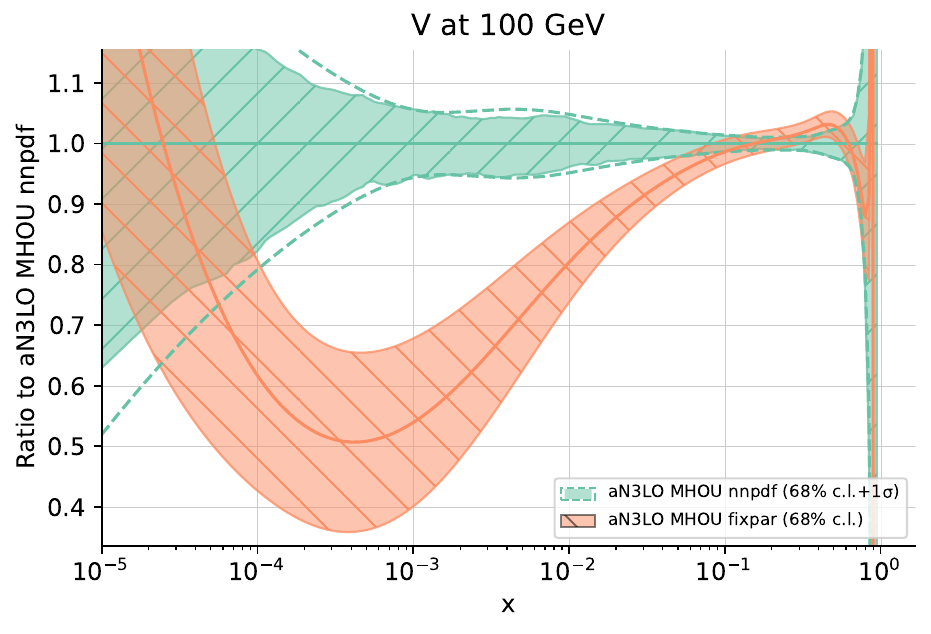}
\includegraphics[scale=0.4]{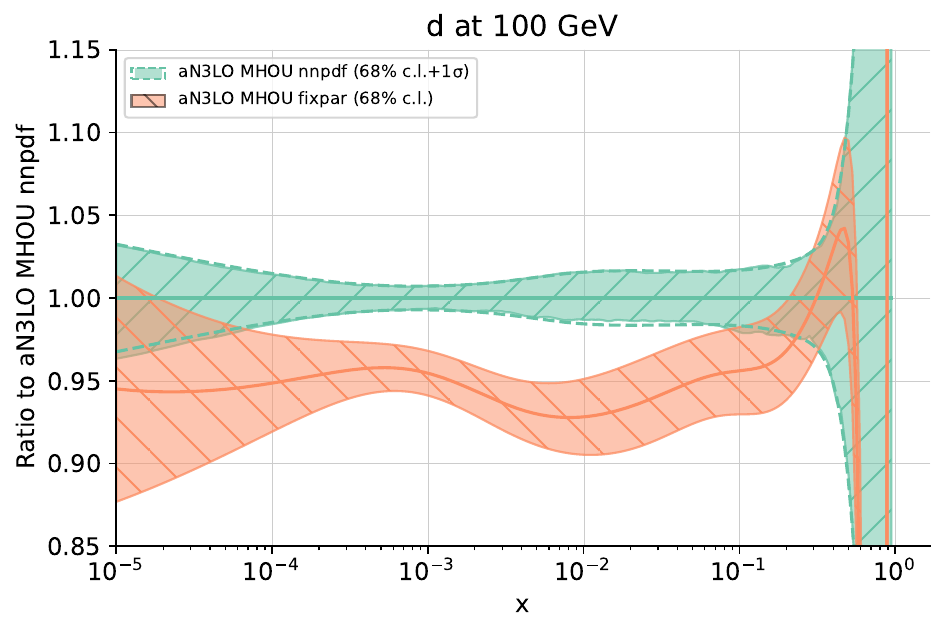}
\end{center}
\caption{\sf Valence (left panel) and down quark (right panel) PDFs found using the fixed parameterisation and NNPDF methodologies. Fixed parameterisation uncertainties are calculated using the dynamic tolerance criterion.}
\label{fig:down_an3lo}
\end{figure}

\paragraph{Valence and down quark PDFs} A non--negligible difference between the two sets of results presented here can be seen for the valence distribution, defined as $V = \sum_i (q_i -\bar{q}_i)$ and plotted in the left panel of Fig.~\ref{fig:down_an3lo}. 
The fixed parameterisation results show a relative dip at medium-$x$ with respect to the NNPDF set.
Since the total integral of the valence distribution is fixed by the sum rules, the medium--$x$ dip is balanced in the low--$x$ extrapolation region by a slower decay to zero of the MSHT results, so that the total integral still gives the correct sum rule for both methodologies.
These kind of effects are likely to be due to some freedom, allowed by the  data entering these fits, in the shape of V, which allows for different implementation of the valence sum rule. It is interesting to note how this is exposed by using two different methodologies, and would not be evident by looking at one single set of results.
This discrepancy is also reflected in differences  at the level of the individual quarks distributions, most notably for the down quark PDF, which is plotted in the right panel of Fig.~\ref{fig:down_an3lo}.

\paragraph{Size of PDF uncertainties}

\begin{figure}
\begin{center}
\includegraphics[scale=0.4]{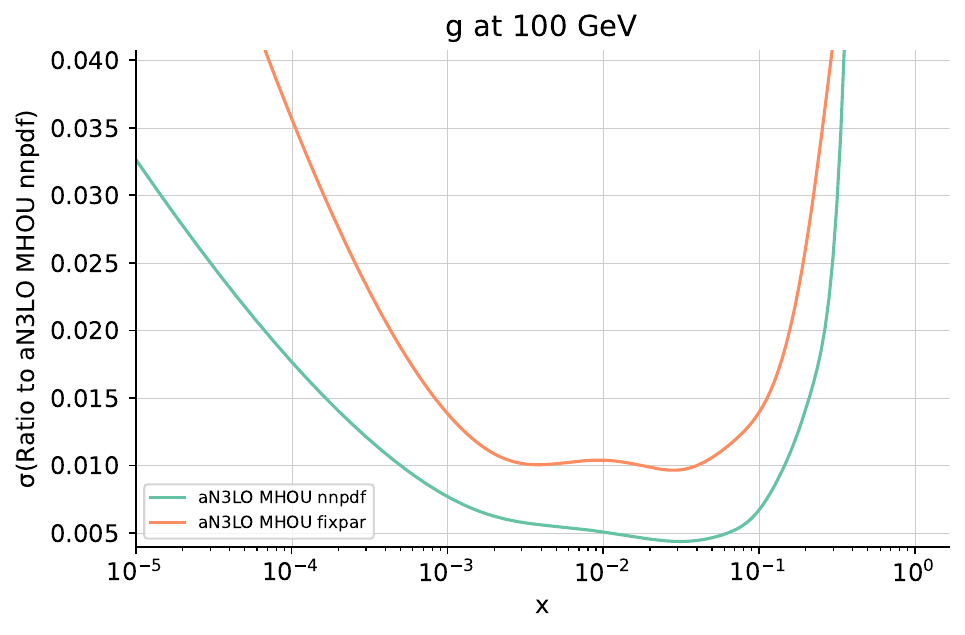}
\includegraphics[scale=0.4]{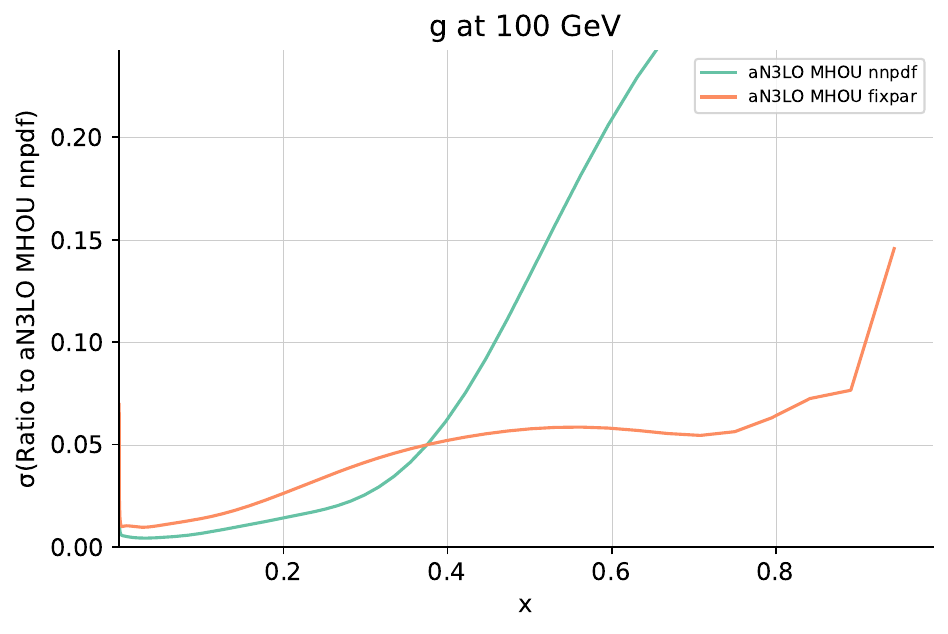}
\end{center}
\caption{\sf Gluon PDF uncertainty found using the fixed parameterisation and NNPDF methodologies. Fixed parameterisation uncertainties are calculated using the dynamic tolerance criterion.}
\label{fig:gluon_unceratinty}
\end{figure}

PDF uncertainties for the gluon PDF at 100 GeV at aN$^3$LO are plotted in Fig.~\ref{fig:gluon_unceratinty}, normalized to the NNPDF central value.
We observe how the NNPDF methodology gives smaller errors in the data and small-$x$ extrapolation region, and bigger uncertainties for the large-$x$ extrapolation region. This trend is not particular to the gluon, and is common to all quark flavors, as is visible for the charm quark PDF shown in Fig.~\ref{fig:gluon_charm_an3lo}. The reasons for these differences has been discussed in detail in Ref.~\cite{Harland-Lang:2024kvt}. In particular, in the data region it is dominantly due to the presence of a tolerance in the fixed parameterisation approach, where as discussed above a dynamic tolerance procedure is applied to calculate the PDF uncertainties. Indeed, as shown in Ref.~\cite{Harland-Lang:2024kvt}, if a fixed $T^2=1$ tolerance is instead used the agreement is generally greatly improved. The difference in the extrapolation region may in large part be driven by the MC replica approach to error propagation applied by NNPDF, which is observed in Ref.~\cite{Harland-Lang:2024kvt} to result in more conservative PDF uncertainties in the high--$x$ region.

In summary, we conclude that the differences previously observed in the aN$^3$LO gluon PDF and in the gluon-gluon luminosity are driven by the specific theory and data settings adopted in the published aN$^3$LO sets, and not by the specific methodological choices. The large-$x$ charm PDFs produced by the two methodologies are also found to be in agreement, despite having different uncertainties. 
Specific features which are instead driven by the fitting methodologies are: the behavior of the valence distribution between $x = 0.5\times10^{-1}$ and $x = 10^{-4}$, resulting in non-negligible tensions in the same $x$ range for individual quark distributions, and the size of the PDF uncertainties. In the latter case, while the fixed parameterisation results have generally more conservative uncertainties in the data and low--$x$ extrapolation region, NNPDF sets have a larger uncertainty in the large-$x$ extrapolation region.

\paragraph{Cross Section Results}

\begin{figure}
\begin{center}
\includegraphics[scale=0.64]{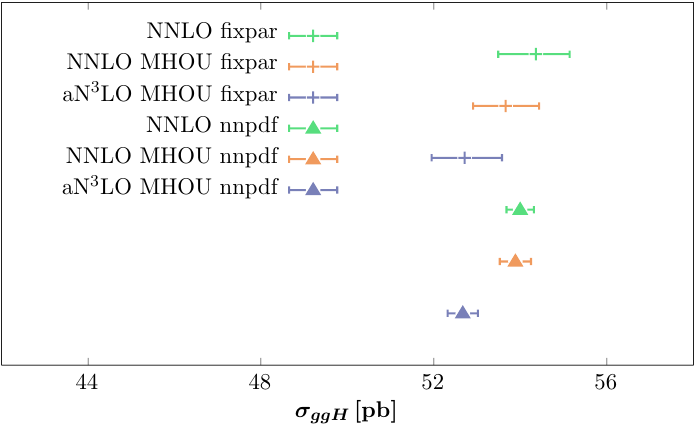}
\includegraphics[scale=0.64]{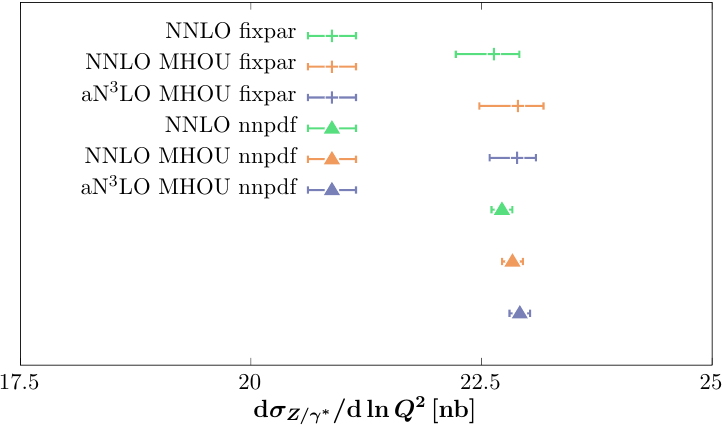}
\includegraphics[scale=0.64]{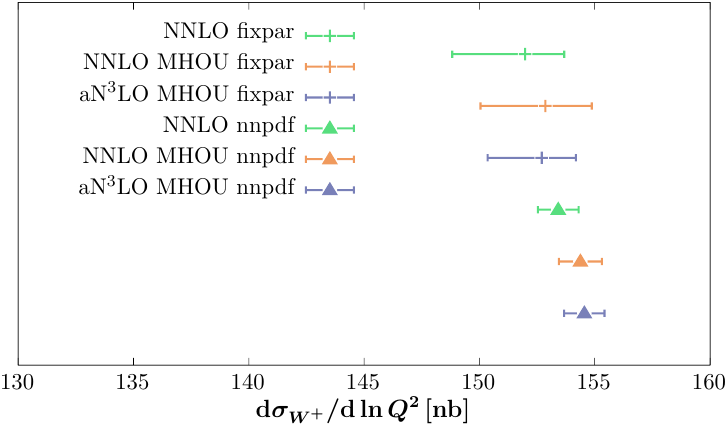}
\includegraphics[scale=0.64]{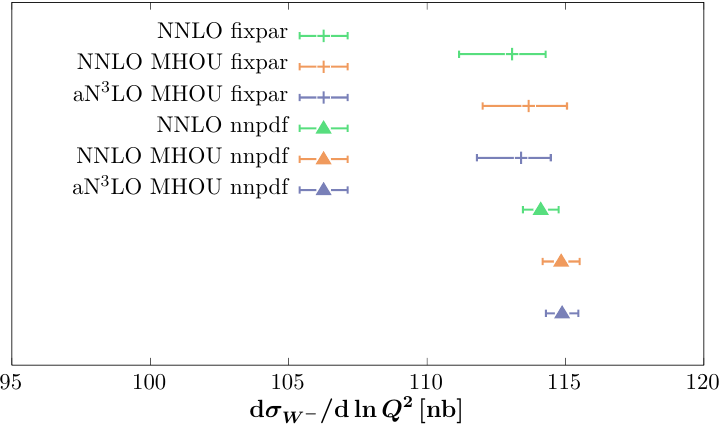}
\end{center}
\caption{\sf Comparison between selected cross section predictions at the $\sqrt{s}=14$ TeV LHC for the NNLO, NNLO+MHOU and aN$^3$LO+MHOU PDFs extracted using the fixed parameterisation and NNPDF methodologies. All cross sections are calculated as described in the text, and correspond to N$^3$LO matrix elements in all cases. PDF uncertainties alone are indicated. }
\label{fig:xs}
\end{figure}

Selected cross section predictions for the NNLO, NNLO+MHOU and aN$^3$LO+MHOU PDFs extracted using the fixed parameterisation and NNPDF methodologies are shown in Fig.~\ref{fig:xs}. These results are calculated using the \texttt{n3loxs} code~\cite{Baglio:2022wzu,Borowka:2018goh,Duhr:2020seh,Duhr:2020sdp,Duhr:2021vwj,Duhr:2019kwi,Duhr:2020kzd,Anastasiou:2014vaa,Mistlberger:2018etf}. In all cases a central scale of $\mu_F=\mu_R=\mu_0$ is taken, with $\mu_0$ corresponding to $m_h/2$, $m_Z$ and $m_W$ for the relevant processes. The $W,Z$ cross sections are evaluated at $Q=M_{W,Z}$. Results are always shown with fixed N$^3$LO matrix elements, to demonstrate the impact of the PDFs alone on the cross sections while PDF uncertainties alone are indicated.

Starting with the Higgs production cross section in gluon fusion, we  can see a trend expected from the luminosity plots in Fig.~\ref{fig:aN3LO_lumi} and previous studies, see Ref.~\cite{Cridge:2024icl}. In particular, there is a moderate but non--negligible reduction in the cross section prediction in going from NNLO to  aN$^3$LO PDFs, while the NNLO+MHOU result is somewhat intermediate. For the $W,Z$ boson predictions the diffference between the  NNLO and  aN$^3$LO PDF results  is somewhat milder, but still noticeable, with the latter PDFs giving a rather larger cross section prediction. The PDF uncertainty on the cross sections is universally larger for the fixed parameterisation case, in line with the results above at the PDF level.

Most importantly, we can clearly see that in all cases the relative trend between the fixed parameterisation and NN results is rather similar. That is, the impact of including aN$^3$LO corrections in the PDF determination on these benchmark cross section predictions is rather insensitive to the methodologoical choice relating to the PDF parameterisation.

\section{Summary and Outlook}\label{sec:conc}

In this paper we have presented a new public framework to perform global fits of PDFs within the Hessian formalism: \fppdf. The general fitting methodology is akin to the one developed by the MSHT collaboration, while the input data and theory computations are taken from the public NNPDF libraries.
The code is open source, and publicly available at
\begin{center}
    \href{https://github.com/FPPDF/fppdf}{https://github.com/FPPDF/fppdf}\ ,
\end{center}
together with runcards and parameters necessary to reproduce the results shown in this paper.
This in particular implies that: the NNPDF implementation of theory computations (hard matrix elements, evolution, matching coefficients) are available at up to aN$^3$LO in QCD, including QED corrections; the full dataset used in the NNPDF4.0 set can be considered, alongside new LHC Run II data which have been implemented since then, see for example Ref.~\cite{Chiefa:2025loi} for a list of some of the new available datasets; theory errors due to missing higher order uncertainties can be included using the theory covariance matrix formalism; the full machinery of closure tests developed within the NNPDF framework can be readily used for validation studies.
By comparing PDFs produced using \fppdf to analogous NNPDF sets, one can achieve a like-for-like comparison, where the only difference between results is given by the methodology for parameterisation the PDFs.
This allows for a study of the impact of different methodological choices on the final PDF central values and, importantly, on the final uncertainties, disentangling completely the impact of different data and theory setting from the methodological driven effects. 

As an example, in this paper we have used \fppdf to perform PDF fits to the same data/theory settings (close to that of NNPDF4.0) but with both NN and fixed PDF parameterisations, at NNLO and aN$^3$LO perturbative orders.
We have then assessed the impact of the two different methodologies on the  central values and uncertainties. We have in particular found that the relative impact of going to the higher perturbative order and/or including missing higher order uncertainties is rather insensitive to which PDF parameterisation is used. This supports the importance of including such corrections, and the fact that the impact of these is due to the genuine effect of including higher--order perturbative corrections in the fit, rather than being tied up with any particular PDF fitting methodology. 

In this paper we have not aimed to explain the differences observed between PDFs produced by different methodologies, but rather to develop a tool to analyse them. An obvious application  of this code, not pursued here, is a detailed study of the size of the PDF uncertainties when comparing  Hessian and Monte Carlo PDFs, investigating for example the role of tolerance, defined in the first but not in the second case, while decoupling all the effects due to data and theory. This study, and others,
such as a detailed benchmarking exercise which will be needed to disentangle the impact of other theoretical settings and those related to data choice.
are left to future work, with the publicly available \fppdf code providing the tool to achieve this.

\paragraph{Acknowledgments} J.C-M. acknowledges funding from the Ramón y Cajal program grant RYC2023-043794-I funded by MCIN/AEI/10.13039/501100011033 and by ESF+. 
TG acknowledges support from the European Union under the MSCA fellowship (Grant Agreement N. 101149419) \textit{Bayesian tools for one-dimensional and three-dimensional hadron structure (Bayhadron)}.
LHL thanks the Science and Technology Facilities Council (STFC) part of U.K.
Research and Innovation for support via the grant award ST/T000856/1. 

\appendix

\section{Code usage and runcard examples}\label{sec:app}

This paper presents the code for \fppdf, an open-source program to perform parametric fit of Parton Distribution Functions (PDFs) using arbitrary functional forms.
Uncertainty estimation is computed as Hessian uncertainties via eigenvector scans (although a replica-based approach can also be taken by repeating the central-value estimation to an ensemble of data replicas).

PDF fits are performed at a fixed scale, using the theory and data made available by the \href{https://github.com/nnpdf/nnpdf}{NNPDF collaboration}~\cite{NNPDF:2021uiq,nnpdfweb}.
This code also uses the NNPDF framework to perform theory predictions and data-theory comparisons required during the optimization.
LHAPDF~\cite{Buckley:2014ana} grids are generated by evolving the fixed-scale PDF with \href{https://github.com/nnpdf/eko}{eko}~\cite{Candido:2022tld}.

\subsection{Installation}

Cloning this repository and installing with pip should install both this code and all necessary dependencies:

\begin{verbatim}
    pip install .
\end{verbatim}

If a python environment is also required, our recommendation is to use \texttt{conda} to create the environment:

\begin{verbatim}
    conda create -n fppdf_environment nnpdf -c conda-forge
    conda activate fppdf_environment
    python -m pip install . --no-deps
\end{verbatim}

this code does not add extra dependencies with respect to \texttt{nnpdf} (and \texttt{nnpdf} is a dependency of this code), therefore this should be enough to get this code and all dependencies in the environment.

After the installation several commands prefixed with \texttt{fppdf\_} will be installed: \texttt{fppdf\_setupfit}, \texttt{fppdf\_fitpdf}, \texttt{fppdf\_hessianerr} and \texttt{fppdf\_evolve}.

\subsection{Running the code}

\subsubsection{Runcard preparation}

The optimization procedure, minimization, and creation of LHAPDF grid is governed by a \texttt{.yaml} runcard, of which there are several examples in this repository (e.g., \texttt{example\_full\_short.yml}).
Below we describe some of the options in these runcards:

\paragraph{Global options}
\begin{itemize}
    \item \texttt{dataset\_inputs}: a list of datasets that will be considered in the fit. It follows the conventions of the NNPDF runcards (see \href{https://github.com/NNPDF/nnpdf/tree/master/n3fit/runcards/examples}{here}) to allow for an easier comparison between our fits and NNPDF fits.
    \item \texttt{posdatasets}: same, for the list of positivity datasets
    \item \texttt{added\_filter\_rules}: extra cuts to add beyond the \href{https://github.com/nnpdf/nnpdf/blob/master/validphys2/src/validphys/cuts/filters.yaml}{default set of cuts} of the NNPDF data 
    \item \texttt{theoryid}: ID defining the theory assumptions under which the optimization is performed. The full database of theory IDs is available \href{https://docs.nnpdf.science/theory/theoryindex.html}{here}
    \item \texttt{genrep}: whether MC replicas should be generated (default \texttt{False}, and should be \texttt{False} for Hessian uncertainties)
    \item \texttt{mcseed}: seed used to generate the MC replicas
\end{itemize}

\paragraph{\texttt{inout\_parameters}}
Parameters to define the input / output of a run
\begin{itemize}
    \item \texttt{inputnam}: file with the starting parameters. Some examples can be seen in the \texttt{input} folder. In these files the second column indicates whether a parameter is free (1) or fixed (0). If they are all fixed no optimization can occur.
    \item \texttt{label}: label of the run, can be arbitrary, output files might have a suffix depending on whether dynamic tolerance is being used
    \item \texttt{covinput}: input covariance matrix file (if used)
    \item \texttt{readcov}: default \texttt{False}, whether to read in a covariance matrix file to do an eigenvector scan.
\end{itemize}

\paragraph{\texttt{fit\_parameters}}
Parameters to change specific options of the optimization.
\begin{itemize}
    \item \texttt{fixpar}: if \texttt{True}, don't perform an optimization, just evaluate the different quantities needed during the fit. Useful for debugging. Default \texttt{False}.
    \item \texttt{nnpdf\_pos}: whether to impose NNPDF positivity in the fit
\end{itemize}

\paragraph{\texttt{chi2\_parameters}}
These parameter control how the uncertainties will be computed
\begin{itemize}
    \item \texttt{dynamic\_tol}: whether to use dynamic tolerance or not
    \item \texttt{t2\_err}: tolerance (squared) to use when \texttt{dyanmic\_tol} is set to \texttt{False}
\end{itemize}

\paragraph{theorycovmatconfig:}
If this namespace exists, the theory covariance matrix will be computed with the following parameters:
\begin{itemize}
    \item \texttt{point\_prescriptions}: The point prescription(s) to be used for the computation of the theory covariance matrix (e.g., 7-points scale variations, ["7 points"])
    \item \texttt{pdf}: Which PDF to use to compute the theory covariance matrix.
    \item \texttt{theoryid}: Theory ID to use for the theory covariance matrix
\end{itemize}

\subsection{Step-by-step fit}

\subsubsection{When including missing higher order uncertainties in the covariance matrix}

Run the following command to prepare the run and compute the theory covariance matrix.
Running this command will download any grids necessary to compute the theory predictions for each of the theories involved in the computation of the covariance matrix and will prepare the theory covariance matrix to be used during optimization.
\begin{verbatim}
fppdf_setupfit <runcard.yml>
\end{verbatim}

\subsubsection{Perform the minimization.}

Performs the fit. If the theory covariance matrix is to be used, \texttt{setupfit} (see above) must be run first.
This procedure might take some hours depending on the number of datasets specified and how far off the minimum the fit started from.
At the end of this procedure the best fit to the central data will be produced.
\begin{verbatim}
fppdf_fitpdf <runcard.yml>
\end{verbatim}

\subsubsection{Compute the Hessian error members.}

Once the fit has finished, used the fit parameters found in the previous step to compute the Hessian error members using the chosen prescription.
\begin{verbatim}
fppdf_Hessianerr <runcard.yml>
\end{verbatim}

\subsubsection{Evolve and create the final grid.}

After the central PDF and the eigenvectors have been computed, they need to be evolve to generate a full LHAPDF grid for all scales. This is obtained with:
\begin{verbatim}
fppdf_evolve <runcard.yml>
\end{verbatim}
which will generate an LHAPDF grid in the \texttt{outputs/evgrids/} folder.

Other output can be found in the \texttt{outputs/} folder:

\begin{itemize}
    \item /evgrids  : the output pdf grids which can then be evolved using the NNPDF evolution (to be tided/documented)
    \item /buffer : outputs from running code
    \item /cov : covariance matrix (used for error calculation)
    \item /evscans: outputs if eigenvector scan is done
    \item /pars : the PDF parameters
\end{itemize}

\bibliography{references}{}
\bibliographystyle{h-physrev}

\end{document}

%% file: tables/tab_chi2_process_msht.tex
\begin{tabularx}{\textwidth}{Xrcccc}
  \toprule
  & &\multicolumn{2}{c}{NNLO (fixed par.)}
  & \multicolumn{2}{c}{aN$^3$LO (fixed par.)}\\
  Dataset
  & $N_{\rm dat}$
  & no MHOU
  & MHOU 
  & no MHOU
  & MHOU \\
  \midrule
  DIS NC
  & 2100 & 1.21 & 1.21
& 1.22 & 1.22 \\
  DIS CC
  &  989 & 0.90 & 0.91 
  & 0.91 & 0.92 \\
  DY NC
  &  735 & 1.21 & 1.18
  & 1.21 & 1.18 \\
  DY CC
  &  157 & 1.43 & 1.36
  & 1.40 & 1.34 \\
  Top pairs
  &   64 & 1.24 & 1.36
  & 1.14 & 1.38 \\
  Single-inclusive jets
  &  356 & 0.90 & 0.76
  & 0.94 & 0.77 \\
  Dijets
  &  144 & 1.94 & 1.61
  & 1.98 & 1.58\\
  Prompt photons 
  &   53 & 0.74 & 0.67
  & 0.75 & 0.68 \\
  Single top
  &   17 & 0.34 & 0.34
  & 0.32 & 0.32 \\
  \midrule
  Total
  & 4616 & 1.15 & 1.13
  & 1.16 & 1.14 \\
\bottomrule
\end{tabularx}

%% file: tables/tab_chi2_process_nnpdf.tex
\begin{tabularx}{\textwidth}{Xrcccc}
  \toprule
  & &\multicolumn{2}{c}{NNLO (NNPDF)}
  & \multicolumn{2}{c}{aN$^3$LO (NNPDF)}\\
  Dataset
  & $N_{\rm \text{dat}}$
  & no MHOU
  & MHOU 
  & no MHOU
  & MHOU \\
  \midrule
  DIS NC
  & 2100 & 1.23 & 1.20
& 1.22 & 1.20 \\
  DIS CC
  &  989 & 0.90 & 0.90 
  & 0.91 & 0.91 \\
  DY NC
  &  735 & 1.19 & 1.15
  & 1.19 & 1.16 \\
  DY CC
  &  157 & 1.46 & 1.34
  & 1.43 & 1.37 \\
  Top pairs
  &   64 & 1.21 & 1.45
  & 1.12 & 1.43 \\
  Single-inclusive jets
  &  356 & 0.96 & 0.80
  & 0.99 & 0.82 \\
  Dijets
  &  144 & 2.01 & 1.73
  & 2.06 & 1.64 \\
  Prompt photons 
  &   53 & 0.74 & 0.68
  & 0.75 & 0.68 \\
  Single top
  &   17 & 0.36 & 0.38
  & 0.35 & 0.36 \\
  \midrule
  Total
  & 4616 & 1.17 & 1.13
  & 1.17 & 1.14 \\
\bottomrule
\end{tabularx}

%% file: main.bbl
\begin{thebibliography}{10}

\bibitem{H1:2015ubc}
H1, ZEUS, H.~Abramowicz {\em et~al.},
\newblock Eur. Phys. J. C {\bf 75}, 580 (2015), 1506.06042.

\bibitem{Alekhin:2017kpj}
S.~Alekhin, J.~Bl\"umlein, S.~Moch, and R.~Placakyte,
\newblock Phys. Rev. D {\bf 96}, 014011 (2017), 1701.05838.

\bibitem{ATLAS:2021vod}
ATLAS, G.~Aad {\em et~al.},
\newblock Eur. Phys. J. C {\bf 82}, 438 (2022), 2112.11266.

\bibitem{Hou:2019efy}
T.-J. Hou {\em et~al.},
\newblock Phys. Rev. D {\bf 103}, 014013 (2021), 1912.10053.

\bibitem{Bailey:2020ooq}
S.~Bailey, T.~Cridge, L.~A. Harland-Lang, A.~D. Martin, and R.~S. Thorne,
\newblock Eur. Phys. J. C {\bf 81}, 341 (2021), 2012.04684.

\bibitem{NNPDF:2021njg}
NNPDF, R.~D. Ball {\em et~al.},
\newblock Eur. Phys. J. C {\bf 82}, 428 (2022), 2109.02653.

\bibitem{PDF4LHCWorkingGroup:2022cjn}
PDF4LHC Working Group, R.~D. Ball {\em et~al.},
\newblock J. Phys. G {\bf 49}, 080501 (2022), 2203.05506.

\bibitem{McGowan:2022nag}
J.~McGowan, T.~Cridge, L.~A. Harland-Lang, and R.~S. Thorne,
\newblock Eur. Phys. J. C {\bf 83}, 185 (2023), 2207.04739.

\bibitem{NNPDF:2024nan}
NNPDF, R.~D. Ball {\em et~al.},
\newblock Eur. Phys. J. C {\bf 84}, 659 (2024), 2402.18635.

\bibitem{CMS:2024ony}
CMS, A.~Hayrapetyan {\em et~al.},
\newblock Phys. Lett. B {\bf 866}, 139526 (2025), 2408.07622.

\bibitem{ATLAS:2023fsi}
ATLAS-CONF-2023-004.

\bibitem{Harland-Lang:2024kvt}
L.~A. Harland-Lang, T.~Cridge, and R.~S. Thorne,
\newblock Eur. Phys. J. C {\bf 85}, 316 (2025), 2407.07944.

\bibitem{Costantini:2025agd}
M.~N. Costantini, L.~Mantani, J.~M. Moore, V.~S. Sanchez, and M.~Ubiali,
\newblock (2025), 2510.03391.

\bibitem{NNPDF:2021uiq}
NNPDF, R.~D. Ball {\em et~al.},
\newblock Eur. Phys. J. C {\bf 81}, 958 (2021), 2109.02671.

\bibitem{Alekhin:2014irh}
S.~Alekhin {\em et~al.},
\newblock Eur. Phys. J. C {\bf 75}, 304 (2015), 1410.4412.

\bibitem{NNPDF:2024dpb}
NNPDF, R.~D. Ball {\em et~al.},
\newblock Eur. Phys. J. C {\bf 84}, 517 (2024), 2401.10319.

\bibitem{levenberg1944method}
K.~Levenberg,
\newblock Quarterly of applied mathematics {\bf 2}, 164 (1944).

\bibitem{DonaldW:2006yco}
M.~Donald~W.,
\newblock J. Soc. Indust. Appl. Math. {\bf 11}, 431 (2006).

\bibitem{Ball:2009qv}
NNPDF, R.~D. Ball {\em et~al.},
\newblock JHEP {\bf 05}, 075 (2010), 0912.2276.

\bibitem{Martin:2009iq}
A.~D. Martin, W.~J. Stirling, R.~S. Thorne, and G.~Watt,
\newblock Eur. Phys. J. C {\bf 63}, 189 (2009), 0901.0002.

\bibitem{Reader:2024cbb}
M.~Reader,
\newblock PoS {\bf DIS2024}, 059 (2025), 2408.12922.

\bibitem{Watt:2012tq}
G.~Watt and R.~S. Thorne,
\newblock JHEP {\bf 08}, 052 (2012), 1205.4024.

\bibitem{Cridge:2024icl}
T.~Cridge {\em et~al.},
\newblock J. Phys. G {\bf 52}, 6 (2025), 2411.05373.

\bibitem{Cridge:2025oel}
T.~Cridge, L.~A. Harland-Lang, and R.~S. Thorne,
\newblock PoS {\bf DIS2025}, 027 (2025), 2510.09321.

\bibitem{Baglio:2022wzu}
J.~Baglio, C.~Duhr, B.~Mistlberger, and R.~Szafron,
\newblock JHEP {\bf 12}, 066 (2022), 2209.06138.

\bibitem{Borowka:2018goh}
S.~Borowka {\em et~al.},
\newblock Comput. Phys. Commun. {\bf 240}, 120 (2019), 1811.11720.

\bibitem{Duhr:2020seh}
C.~Duhr, F.~Dulat, and B.~Mistlberger,
\newblock Phys. Rev. Lett. {\bf 125}, 172001 (2020), 2001.07717.

\bibitem{Duhr:2020sdp}
C.~Duhr, F.~Dulat, and B.~Mistlberger,
\newblock JHEP {\bf 11}, 143 (2020), 2007.13313.

\bibitem{Duhr:2021vwj}
C.~Duhr and B.~Mistlberger,
\newblock JHEP {\bf 03}, 116 (2022), 2111.10379.

\bibitem{Duhr:2019kwi}
C.~Duhr, F.~Dulat, and B.~Mistlberger,
\newblock Phys. Rev. Lett. {\bf 125}, 051804 (2020), 1904.09990.

\bibitem{Duhr:2020kzd}
C.~Duhr, F.~Dulat, V.~Hirschi, and B.~Mistlberger,
\newblock JHEP {\bf 08}, 017 (2020), 2004.04752.

\bibitem{Anastasiou:2014vaa}
C.~Anastasiou {\em et~al.},
\newblock Phys. Lett. B {\bf 737}, 325 (2014), 1403.4616.

\bibitem{Mistlberger:2018etf}
B.~Mistlberger,
\newblock JHEP {\bf 05}, 028 (2018), 1802.00833.

\bibitem{Chiefa:2025loi}
A.~Chiefa {\em et~al.},
\newblock JHEP {\bf 07}, 067 (2025), 2501.10359.

\bibitem{nnpdfweb}
\texttt{https://docs.nnpdf.science}.

\bibitem{Buckley:2014ana}
A.~Buckley {\em et~al.},
\newblock Eur. Phys. J. C {\bf 75}, 132 (2015), 1412.7420.

\bibitem{Candido:2022tld}
A.~Candido, F.~Hekhorn, and G.~Magni,
\newblock Eur. Phys. J. C {\bf 82}, 976 (2022), 2202.02338.

\end{thebibliography}
